\documentclass[11pt,preprint,showpacs,preprintnumbers,amsmath,amssymb]{revtex4}

\usepackage{graphicx}
\usepackage{dcolumn}
\usepackage{bm}

\usepackage{CJK}

\begin{document}

\preprint{Physical Review D 92, 076008 (2015)}

\begin{CJK*}{GBK}{}

\title{ Masses of doubly and triply charmed baryons }

\author{Ke-Wei Wei\footnote{e-mail: weikw@hotmail.com}}
\affiliation{\scriptsize{College of Physics and Electrical Engineering, Anyang Normal University, Anyang 455000, China}}

\author{Bing Chen\footnote{e-mail: chenbing@shu.edu.cn}}
\affiliation{\scriptsize{College of Physics and Electrical Engineering, Anyang Normal University, Anyang 455000, China}}

\author{Xin-Heng Guo\footnote{Corresponding author, e-mail: xhguo@bnu.edu.cn}}
\affiliation{\scriptsize{College of Nuclear Science and Technology, Beijing Normal University, Beijing 100875, China}}

\date{\today\\}

\begin{abstract}
    Until now, the first reported doubly charmed baryon $\Xi_{cc}^{+}(3520)$ is still  a puzzle. It was discovered and confirmed by SELEX collaboration, but not confirmed by LHCb,  BABAR, BELLE, FOCUS, or any other collaboration.
    In the present paper, by employing Regge phenomenology, we first express the  mass  of  the ground state ($L$=0) doubly charmed baryon  $\Omega_{cc}^{*+}$
 as a function of masses of the well established light baryons and singly charmed baryons.
    Inserting the recent experimental data, the mass of $\Omega_{cc}^{*+}$ is given to be 3809$\pm$36 MeV,  which is independent of any unobservable parameters.   
    Then, with the quadratic mass relations, we calculate the  masses of  the ground state triply charmed baryon $\Omega_{ccc}^{++}$ and    doubly charmed baryons $\Xi_{cc}^{(*)++}$, $\Xi_{cc}^{(*)+}$, and $\Omega_{cc}^{+}$
(the mass of  $\Xi_{cc}^{+}$ is determined as 3520$^{+41}_{-40}$ MeV, which agrees with the mass of $\Xi_{cc}^{+}(3520)$).    
    The isospin splitting $M_{\Xi_{cc}^{++}} - M_{\Xi_{cc}^{+}} = 0.4 \pm 0.3$ MeV.
 After that, masses of the orbitally excited ($L$=1,2,3) doubly and triply charmed baryons are estimated.    
     The results  are reasonable    comparing with those extracted in many other approaches.   
    We suggest more efforts to study doubly and triply charmed baryons both theoretically and experimentally, not only for the abundance of baryon spectra,
 but also for numerically examining whether the linear mass relations or the quadratic mass relations are realized in nature.   
 Our  predictions are useful for the discovery of unobserved doubly and triply charmed baryon states and the $J^P$ assignment of these states.

\end{abstract}


\maketitle
\end{CJK*}

\section{Introduction}
    In Quark Model,  doubly and triply charmed baryons exist \cite{PDG2014}.
    Many theoretical works have been focused on the mass spectra of doubly and triply charmed baryons in different approaches \cite{Regge2008,
    Burakovsky1997, Ponce-1979, Vijande-etal-2004, Martynenko:2007je, {Chiu:2005zc}, H.R.Petry, Roberts:2007ni, Bjorken:1985ei, Flynn-etal-2003, Roncaglia:1995az, Ebert:2002-05-2007, He:2004px, Kiselev-etal-2000-2002,
 Martin-Richard-1995, Kiselev:2000jb, Jia-2006, Hasenfratz:1980ka,    Valcarce:2008dr,
    Brown:2014ena, Brown:2014pra, Karliner:2014gca, Ghalenovi:2014swa, Aliev:2012iv, Briceno:2012wt, Wang:2010vn, Gerasyuta:1999pc, Weng:2010rb, Alexandrou:2012xk, Alexandrou:2014sha, Zhang:2008rt, Kiselev:1999zj,
 Bagan:1992za, Patel:2008mv, Bernotas:2008bu, Ghalenovi:2011zz, Jaffe:1975us, Kiselev:2001fw, Tang:2011fv, Giannuzzi:2009gh,   Aliev:2014lxa, Padmanath:2013zfa, LlanesEstrada:2011kc, Flynn:2011gf}.
%
    Experimentally, the doubly charmed baryon $\Xi_{cc}^+(3520)$ (ccd) was first reported in the charged decay mode $\Xi_{cc}^+ \rightarrow \Lambda_c^+ K^- \pi^+$ (SELEX 2002~\cite{Xi_cc-SELEX2002})
 and confirmed in another charged decay mode $\Xi_{cc}^+ \rightarrow p D^+ K^-$ (SELEX 2005)~\cite{Xi_cc-SELEX2005}.
   However, the $J^P$ number of $\Xi_{cc}^+(3520)$ has not been determined experimentally.
    Until recently, $\Xi_{cc}^+(3520)$ was the only doubly charmed state adopted by \emph{Particle Data Group} with the average mass 3518.9$\pm$0.9 MeV \cite{PDG2014}.
    In our previous work in 2008 \cite{Regge2008}, it was illustrated that if the state $\Xi_{cc}^+(3520)$ really exists,
 the mass of $\Xi_{cc}^+(3520)$ is too small to be assigned as the $\frac{3}{2}^+$ doubly charmed baryons, and its spin-parity should be $J^P=\frac{1}{2}^+$.
    Under this assumption,  masses of other doubly and triply charmed baryons (${\Omega_{cc}}$, ${\Xi_{cc}^*}$, ${\Omega_{cc}^*}$, and ${\Omega_{ccc}}$) were  calculated in Ref. \cite{Regge2008}.
    However, curiously, until now, $\Xi_{cc}^+(3520)$ has not been confirmed by any other collaboration (notably by LHCb \cite{Xi_cc(3520)-LHCb}, BELLE \cite{Xi_cc(3520)-BELLE}, BABAR \cite{Xi_cc(3520)-BABAR}, and FOCUS \cite{Xi_cc(3520)-FOCUS}),
 even though they have more reconstructed charm baryons than SELEX. 
    Thus, the existence of  $\Xi_{cc}^+(3520)$ is doubtful.  
    Therefore, the way to calculate the masses of the  doubly and triply charmed baryons which is independent of $\Xi_{cc}^+(3520)$ is necessary.
    According to the latest ``Review of Particle Physics'' (RPP) \cite{PDG2014},
 many light baryons and singly charmed baryons have been well established.  They are labeled with three or four stars in Baryon Summary Table
  while $\Xi_{cc}^+(3520)$ is just labeled  with one star.
%
%
%
%
    Therefore, we will  attempt to avoid using the mass of $\Xi_{cc}^+(3520)$   and focus on searching mass relations which can express the mass of a doubly charmed baryon as a function of masses of the well established light baryons and singly charmed baryons.
 This is the most motivation of this work.
    It is noted that the direct generalization of the Gell-Mann-Okubo formula \cite{Gell-Mann-Okubo} to the charmed and bottom hadrons cannot agree well with experimental data.   
    The Regge trajectory ansatz is a simple and effective phenomenological model to study mass spectra \cite{De-minLi2004, Zhang:2008rt, Wei:2010zza, Ailin Zhang, non-linear, anisovich, Ebert:2011kk, Masjuan:2012gc}
 and mass relations \cite{Regge2008,  Burakovsky-meson-relation, Burakovsky1997, Kaidalov1982} for baryons and mesons.

   In the present paper, in Regge phenomenology, we will first express the  mass  of  the ground state (the orbital quantum number $L=0$) doubly charmed baryon  $\Omega_{cc}^{*+}$
 as a function of masses of light baryons ($\Sigma^{*+}$, $\Xi^{*0}$, $\Omega^-$) and singly charmed baryons ($\Xi_c^{*+}$, $\Omega_{c}^{*0}$) which are well established experimentally.
    With the help of the relation between the slope ratio and masses of baryons with different flavors, we will extract the values of slopes for doubly and triply charmed baryons.
    Then, with the quadratic mass relations derived from Regge phenomenology, we will calculate the  masses of  the ground state triply charmed baryon $\Omega_{ccc}^{++}$ and  doubly charmed baryons $\Xi_{cc}^{(*)++}$, $\Xi_{cc}^{(*)+}$, and $\Omega_{cc}^{+}$.
 After that, masses of the orbitally excited ($L$=1,2,3) doubly and triply charmed baryons will be estimated.    
     The results will be compared with the existing experimental data and those suggested in many other approaches \cite{Regge2008,
    Burakovsky1997, Ponce-1979, Vijande-etal-2004, Martynenko:2007je, {Chiu:2005zc}, H.R.Petry, Roberts:2007ni, Bjorken:1985ei, Flynn-etal-2003, Roncaglia:1995az, Ebert:2002-05-2007, He:2004px, Kiselev-etal-2000-2002,
 Martin-Richard-1995, Kiselev:2000jb, Jia-2006, Hasenfratz:1980ka,       Valcarce:2008dr,
    Brown:2014ena, Brown:2014pra, Karliner:2014gca, Ghalenovi:2014swa, Aliev:2012iv, Briceno:2012wt, Wang:2010vn, Gerasyuta:1999pc, Weng:2010rb, Alexandrou:2012xk, Alexandrou:2014sha, Zhang:2008rt, Kiselev:1999zj,
 Bagan:1992za, Patel:2008mv, Bernotas:2008bu, Ghalenovi:2011zz, Jaffe:1975us, Kiselev:2001fw, Tang:2011fv, Giannuzzi:2009gh,   Aliev:2014lxa, Padmanath:2013zfa, LlanesEstrada:2011kc, Flynn:2011gf}.

    The remainder of this paper is organized as follows.
    In Sec. II, we  first briefly introduce the Regge trajectory ansatz.
 Then, we express the quadratic mass  of doubly charmed baryon $\Omega_{cc}^{*+}$  as a function of masses of light baryons  and singly charmed baryons.
 After that, we  calculate the slopes of doubly and triply charmed baryons and the masses of the doubly charmed baryons  lying on the $\Omega_{cc}^{*+}$ trajectory.
    In Sec. III, 
 the masses of baryons lying on the  ${\Omega_{ccc}^{++}}$ and ${\Xi_{cc}^{*}}$  trajectories are extracted.
    In Sec. IV,  
 the masses of baryons lying on the ${\Xi_{cc}^{}}$ and ${\Omega_{cc}^{+}}$ trajectories are estimated.
    In Sec. V, we give a discussion and summary.

\section{Mass of the doubly charmed baryon $\Omega_{cc}^{*+}$}

  Regge theory  is concerned with almost all aspects of strong interactions, including particle spectra,   forces between particles, and the high energy behavior of scattering amplitudes \cite{regge-book}.
    It is known from Regge theory that all mesons and baryons are associated with Regge trajectories
 (Regge poles which move in the complex angular momentum plane as a function of energy)~\cite{Regge-theory}.
  Hadrons lying on the same Regge trajectory which have the same internal quantum numbers are classified into the same family \cite{Chew-Frautschi,regge-book},
\begin{equation}
J=\alpha(M)=a(0)+\alpha^\prime M^2, \label{regge1}
\end{equation}
 where $\alpha^\prime$ and $a(0)$ are respectively the intercept and slope of the Regge trajectory on which the particles lie.
     For a baryon multiplet, the Regge intercepts and Regge slopes for different flavors can be related by the following relations (see Ref.~\cite{Regge2008} and references therein):
\begin{equation}
a_{iik}(0)+a_{jjk}(0)=2a_{ijk}(0) \label{intercept2},
\end{equation}
\begin{equation}
\frac{1}{\alpha_{iik}^\prime}+\frac{1}{\alpha_{jjk}^\prime}=\frac{2}{\alpha_{ijk}^\prime}, \label{slope1b}
\end{equation}
where $i$, $j$, and $k$ represent quark flavors.
    There is also a relation  about the factorization of slopes, 
 ${\alpha_{ijq}^\prime}^2={\alpha_{iiq}^\prime} \times {\alpha_{jjq}^\prime}$ \cite{residues-baryon}.
    This relation is consistent with the formal chiral limit, but fails in the heavy quark limit \cite{Burakovsky-para-relation}.
 Charm quark is one of the  heavy quarks. Therefore, we do not use this relation  about the factorization of slopes in the present paper.

      Using Eqs. (\ref{regge1}), (\ref{intercept2}), and (\ref{slope1b}),  when the quark masses $m_j > m_i$, one can obtain

\begin{equation}
 \frac{\alpha_{jjk}'}{\alpha_{iik}'} = \frac{1}{2M_{jjk}^2}\times[(4M_{ijk}^2-M_{iik}^2-M_{jjk}^2)+\sqrt{(4M_{ijk}^2-M_{iik}^2-M_{jjk}^2)^2-4M_{iik}^2M_{jjk}^2} ]. \label{solution-b}
\end{equation}
    Therefore, for the $\frac{3}{2}^+$ multiplet,
\begin{equation}
\begin{aligned}
\frac{\alpha_{scc}'}{\alpha_{uus}'} =& \frac{1}{2M_{\Omega_{cc}^{*+}}^2}\times[(4M_{\Xi_c^{*+}}^2-M_{\Sigma^{*+}}^2-M_{\Omega_{cc}^{*+}}^2)+\sqrt{(4M_{\Xi_c^{*+}}^2-M_{\Sigma^{*+}}^2-M_{\Omega_{cc}^{*+}}^2)^2-4M_{\Sigma^{*+}}^2 M_{\Omega_{cc}^{*+}}^2} ],  \\
\frac{\alpha_{sss}'}{\alpha_{uus}'} &= \frac{1}{2M_{\Omega^-}^2}\times[(4M_{\Xi^{*0}}^2-M_{\Sigma^{*+}}^2-M_{\Omega^-}^2)+\sqrt{(4M_{\Xi^{*0}}^2-M_{\Sigma^{*+}}^2-M_{\Omega^-}^2)^2-4M_{\Sigma^{*+}}^2 M_{\Omega^-}^2} ],  \\
\frac{\alpha_{scc}'}{\alpha_{sss}'} &= \frac{1}{2M_{\Omega_{cc}^{*+}}^2}\times[(4M_{\Omega_{c}^{*0}}^2-M_{\Omega^-}^2-M_{\Omega_{cc}^{*+}}^2)+\sqrt{(4M_{\Omega_{c}^{*0}}^2-M_{\Omega^-}^2-M_{\Omega_{cc}^{*+}}^2)^2-4M_{\Omega^-}^2M_{\Omega_{cc}^{*+}}^2} ]. \label{ccs}
\end{aligned}
\end{equation}
    From Eq. (\ref{ccs}), we can obtain the squared mass of $\Omega_{cc}^{*+}$ as a function of the squared masses of $\Sigma^{*+}$, $\Xi^{*0}$, $\Omega^-$, $\Xi_c^{*+}$ and $\Omega_{c}^{*0}$,
%
{\tiny
\begin{equation}
\begin{aligned}
& M^2_{\Omega_{cc}^{*+}}=
\frac{2 M_{\Omega_{c}^{*0}}^2 \left(M_{\Omega^-}^2-4 M_{\Xi^{*0}}^2+3 M_{\Sigma^{*+}}^2\right) -M_{\Omega^-}^2 \left(M_{\Omega^-}^2-6 M_{\Xi_c^{*+}}^2-6 M_{\Xi^{*0}}^2+2 M_{\Sigma^{*+}}^2\right) -\left(4 M_{\Xi^{*0}}^2-M_{\Sigma^{*+}}^2\right) \left(2 M_{\Xi_c^{*+}}^2+2 M_{\Xi^{*0}}^2-M_{\Sigma^{*+}}^2\right)}{2 \left(M_{\Omega^-}^2-2 M_{\Xi^{*0}}^2+M_{\Sigma^{*+}}^2\right)} +
\\
& \frac{\sqrt{M_{\Omega^-}^4-2 M_{\Omega^-}^2 \left(4 M_{\Xi^{*0}}^2+M_{\Sigma^{*+}}^2\right)+\left(M_{\Sigma^{*+}}^2-4 M_{\Xi^{*0}}^2\right){}^2} \sqrt{\left(M_{\Omega^-}^2-2 M_{\Xi_c^{*+}}^2-2 M_{\Xi^{*0}}^2+M_{\Sigma^{*+}}^2\right)^2 -4 M_{\Omega_{c}^{*0}}^2 \left(M_{\Omega^-}^2+2 M_{\Xi_c^{*+}}^2-2 M_{\Xi^{*0}}^2+M_{\Sigma^{*+}}^2 -M_{\Omega_{c}^{*0}}^2\right)}}{2 \left(M_{\Omega^-}^2-2 M_{\Xi^{*0}}^2+M_{\Sigma^{*+}}^2\right)}
\\
& ~~~~ ~~~~=
\frac{1}{2} \left(2 M_{\Omega_{c}^{*0}}^2+6 M_{\Xi_c^{*+}}^2+4 M_{\Xi^{*0}}^2-M_{\Sigma^{*+}}^2\right)-\frac{M_{\Omega^-}^2}{2}-\frac{2 \left(M_{\Omega_{c}^{*0}}^2-M_{\Xi_c^{*+}}^2\right) \left(M_{\Xi^{*0}}^2-M_{\Sigma^{*+}}^2\right)}{M_{\Omega^-}^2-2 M_{\Xi^{*0}}^2+M_{\Sigma^{*+}}^2}+
\\
& \frac{\sqrt{M_{\Omega^-}^4-2 M_{\Omega^-}^2 \left(4 M_{\Xi^{*0}}^2+M_{\Sigma^{*+}}^2\right)+\left(M_{\Sigma^{*+}}^2-4 M_{\Xi^{*0}}^2\right){}^2} \sqrt{\left(M_{\Omega^-}^2-2 M_{\Xi_c^{*+}}^2-2 M_{\Xi^{*0}}^2+M_{\Sigma^{*+}}^2\right)^2 -4 M_{\Omega_{c}^{*0}}^2 \left(M_{\Omega^-}^2+2 M_{\Xi_c^{*+}}^2-2 M_{\Xi^{*0}}^2+M_{\Sigma^{*+}}^2 -M_{\Omega_{c}^{*0}}^2\right)}}{2 \left(M_{\Omega^-}^2-2 M_{\Xi^{*0}}^2+M_{\Sigma^{*+}}^2\right)}. \label{ccs-express}
\end{aligned}
\end{equation}
}

    From the latest RPP \cite{PDG2014}, $M_{\Sigma^{*+}} = 1382.80 \pm 0.35$ MeV,  $M_{\Xi^{*0}} = 1531.80 \pm 0.32$ MeV,  $M_{\Omega^-} = 1672.45 \pm 0.29$ MeV,  $M_{\Xi_c^{*+}} = 2645.9 ^{+0.5}_{-0.6}$ MeV,  $M_{\Omega_{c}^{*0}} = 2765.9 \pm 2.0$ MeV.
     Inserting these mass values into the relation (\ref{ccs-express}),
 one can get $M_{\Omega_{cc}^{\ast+}}$ = 3809$\pm$36 MeV (where the uncertainty comes from the measurement errors of the  input baryons).  
%
    In this work, all the masses of baryons used in the calculation are taken from RPP~\cite{PDG2014} and the mass differences between isospin multiplet are also considered.
    Comparison of the masses of $\Omega_{cc}^{*+}$ extracted in the present paper and those given in other references is shown in Table 1.



    From Eq. (\ref{ccs}),
 with the help of  $\alpha_{uus}'=2/(M_{\Sigma(2030)}^{2}-M_{\Sigma^{*++}}^{2})$,   
 one can have  the expression for $\alpha_{scc}'$ and get its value ($\alpha_{scc}'=0.509^{+0.053}_{-0.054}$ GeV$^{-2}$).
%
%
    Similarly, using the masses of baryons presented in Eq. (\ref{ccs}), with the aid of Eq. (\ref{slope1b}), one can have the expressions for  $\alpha^\prime_{ccc}$ and $\alpha^\prime_{ucc}$ and get their values
%
 ($\alpha^\prime_{ccc}$ = 0.423$^{+0.055}_{-0.056}$ GeV$^{-2}$,
 $\alpha^\prime_{ucc}$ = 0.516$^{+0.056}_{-0.057}$ GeV$^{-2}$).
%
%

   From Eq. (\ref{regge1}), one has
{\small
\begin{equation}
M_{J+1}=\sqrt{M_J^2+\frac{1}{\alpha'}}  . \label{M-J+1}
\end{equation}
}
   Then, with the expressions for $M_{\Omega_{cc}^{\ast +}}$ and $\alpha_{scc}'$ obtained above, from Eq. (\ref{M-J+1}),
 the masses of the orbitally excited baryons ($L$=1,2,3, while $J^P=\frac{5}{2}^-,\frac{7}{2}^+,\frac{9}{2}^-$) lying on the $\Omega_{cc}^{*+}$ trajectory
 can be expressed as functions of masses of light baryons and singly charmed baryons.
  The numerical results are also shown in Table 1.

   \section{Masses of the baryons lying on the $\Omega_{ccc}^{++}$ and $\Xi_{cc}^{*}$ trajectories}

    Based on Eqs.  (\ref{intercept2}) and (\ref{slope1b}), we can introduce two parameters $\gamma_{x}$ and  $\lambda_{x}$,
    $\gamma_{x} = \frac{1}{\alpha_{uux}^\prime}-\frac{1}{\alpha_{uuu}^\prime}$, $\lambda_{x} = a_{uuu}(0)-a_{uux}(0)$ ($x$ denotes the flavor of a quark).
     To evaluate the high-order effects, we  introduce the parameter $\delta$,  
{\small
\begin{equation}
\delta_{ij,k} \equiv M_{iik}^2+M_{jjk}^2-2M_{ijk}^2.
\end{equation}
}
%
%
%
  Combining with   Eq. (\ref{regge1}), we can have
{\small
\begin{equation}
\begin{split}
\delta_{ij,k} = M_{iik}^2+M_{jjk}^2-2M_{ijk}^2
              = 2(\lambda_i-\lambda_j)(\gamma_i-\gamma_j) \label{baryon-equal} .
\end{split}
\end{equation}
}
 From Eq. (\ref{baryon-equal}), we can  see that $\delta_{ij,k}$ is independent of the  $k$ quark.



  For the $\frac{3}{2}^+$ multiplet,
 noticing that $\delta_{ij,k}^{\frac{3}{2}^+}$ in the above relation (\ref{baryon-equal}) is independent of $k$,  considering the difference of u-quark and d-quark,
   when $i=s$, $j=c$, $k=u,d,s,c$, Eq. (\ref{baryon-equal}) can be expressed as follow:
{\small 
\begin{equation}
\delta_{sc}^{\frac{3}{2}^{+}}
=M_{\Xi^{\ast 0}}^2+M_{\Xi_{cc}^{\ast ++}}^2-2M_{\Xi_c^{\ast +}}^2
=M_{\Xi^{\ast -}}^2+M_{\Xi_{cc}^{\ast +}}^2-2M_{\Xi_c^{\ast 0}}^2
=M_{\Omega^{-}}^2+M_{\Omega_{cc}^{\ast +}}^2-2M_{\Omega_c^{\ast 0}}^2=M_{\Omega_c^{\ast 0}}^2+M_{\Omega_{ccc}^{++}}^2-2M_{\Omega_{cc}^{\ast +}}^2
 \label{subeq:3sc-3+} .
\end{equation}
}
%
    With the expression for the mass of $\Omega_{cc}^{*+}$ (Eq.(\ref{ccs-express})), from the quadratic mass equations  Eq. (\ref{subeq:3sc-3+}),
 we can obtain the expressions for the masses of $\Xi_{cc}^{*++}$, $\Xi_{cc}^{*+}$,  and $\Omega_{ccc}^{++}$.
  For example,
%
{\tiny
\begin{equation}
\begin{aligned}
& M^2_{\Omega_{ccc}^{++}}=
\frac{ 2 M_{\Omega^-}^2 \left(9 M_{\Xi_c^{*+}}^2+7 M_{\Xi^{*0}}^2-2 M_{\Sigma^{*+}}^2\right) -M_{\Omega^-}^4 -3 \left(4 M_{\Omega_c^{*0}}^2 \left(M_{\Xi^{*0}}^2-M_{\Sigma^{*+}}^2\right)+\left(4 M_{\Xi^{*0}}^2-M_{\Sigma^{*+}}^2\right) \left(2 M_{\Xi_c^{*+}}^2+2 M_{\Xi^{*0}}^2-M_{\Sigma^{*+}}^2\right)\right)}{2 \left(M_{\Omega^-}^2-2 M_{\Xi^{*0}}^2+M_{\Sigma^{*+}}^2\right)} +
\\
& \frac{3 \sqrt{ M_{\Omega^-}^4+\left(M_{\Sigma^{*+}}^2-4 M_{\Xi^{*0}}^2\right)^2 -2 M_{\Omega^-}^2 \left(4 M_{\Xi^{*0}}^2+M_{\Sigma^{*+}}^2\right)} \sqrt{4 M_{\Omega_c^{*0}}^4+\left(M_{\Omega^-}^2-2 M_{\Xi_c^{*+}}^2-2 M_{\Xi^{*0}}^2+M_{\Sigma^{*+}}^2\right)^2 -4 M_{\Omega_c^{*0}}^2 \left(M_{\Omega^-}^2+2 M_{\Xi_c^{*+}}^2-2 M_{\Xi^{*0}}^2+M_{\Sigma^{*+}}^2\right)}}{2 \left(M_{\Omega^-}^2-2 M_{\Xi^{*0}}^2+M_{\Sigma^{*+}}^2\right)}
. \label{ccc-express}
\end{aligned}
\end{equation}
}
%
%
    Inserting the masses \cite{PDG2014} of $\Sigma^{*+}$, $\Xi^{*0}$, $\Omega^-$, $\Xi_c^{*+}$,   and $\Omega_c^{*0}$    
     into Eq. (\ref{ccc-express}), one has $M_{\Omega_{ccc}^{++}}$=4834$^{+82}_{-81}$ MeV (truncated to the 1 MeV digit).
    Similarly, we can get $M_{\Xi_{cc}^{*++}}$=3696$\pm33$ MeV, $M_{\Xi_{cc}^{*+}}$=3695$\pm35$ MeV.
  We use $M_{\Xi_{cc}^{*}}$ to note the averaged mass of  ${\Xi_{cc}^{*++}}$ and ${\Xi_{cc}^{*+}}$.
  Therefore, we can obtain the expression for $M_{\Xi_{cc}^{*}}$ and get its  value to be 3695$\pm$34 MeV.
    We can also obtain the expression for the  isospin splitting $M_{\Xi_{cc}^{*++}} - M_{\Xi_{cc}^{*+}}$  and get its  value to be $1.3^{+1.1}_{-1.2}$ MeV.
 We can keep one digit after the decimal point in this result  because  all the input data in the expression  have one or more digits after the decimal point.
    The masses of $\Omega_{ccc}^{}$  and $\Xi_{cc}^{*}$ extracted in the present paper and those given in other references are shown in Table 2 and Table 3, respectively.
%


    Then, with the mass expressions for $\Omega_{ccc}^{++}$, $\Xi_{cc}^{*}$, $\alpha^\prime_{ccc}$ and $\alpha^\prime_{ucc}$ obtained above,
   the masses of   the baryons lying on the $\Omega_{ccc}^{++}$ and $\Xi_{cc}^{*}$ trajectories  can be expressed as functions of masses of light baryons and singly charmed baryons.
    The numerical results are also shown in Tables 2 and 3, respectively.
    The wave function of a baryon  is antisymmetry.
    According to the quantum number analysis,  the odd-parity  $\Omega_{ccc}^{++}$ baryons cannot have the total quark spin $S$=$3\over{2}$.
  Therefore, the  odd-parity    $\Omega_{ccc}^{++}$ baryons ($L$=1,3) have the spin-parities $J^P$=$\frac{3}{2}^-$, $\frac{7}{2}^-$.
    Our calculation indicates that the isospin splittings for the orbital excited states are very small.
   Therefore, we only report the averaged mass of the two isospin partner  in Table 3.


\section{Masses of the   baryons lying on the $\Xi_{cc}$ and $\Omega_{cc}^{}$ trajectories}

    For the $\frac{1}{2}^+$ multiplet, $\delta_{nc}^{\frac{1}{2}^+}$ can be expressed as ($n$ denotes the quark $u$ or $d$)
\begin{equation}
\delta_{uc,d}^{\frac{1}{2}^{+}} + \delta_{dc,u}^{\frac{1}{2}^{+}}
= M^2_{N^+}+M^2_{\Xi_{cc}^+} - 2(\frac{3M_{\Lambda_c^+}^2+M_{\Sigma_c^+}^2}{4})
+ M^2_{N^0}+M^2_{\Xi_{cc}^{++}} - 2(\frac{3M_{\Lambda_c^+}^2+M_{\Sigma_c^+}^2}{4})
 \label{subeq:uc+dc-1+} .
\end{equation}
%
%
     Based on Eq. (\ref{baryon-equal}), when $i=u(d)$, $j=c$, $k=d(u),s$, $\delta_{nc}^{\frac{3}{2}^+}$ can be expressed as
\begin{equation}
\begin{aligned}
\delta_{uc,d}^{\frac{3}{2}^{+}}
=\delta_{uc,s}^{\frac{3}{2}^{+}}
=M_{\Sigma^{\ast +}}^2+M_{\Omega_{cc}^{\ast +}}^2-2M_{\Xi_c^{\ast +}}^2 ,
%
\\
%
\delta_{dc,u}^{\frac{3}{2}^{+}}
=\delta_{dc,s}^{\frac{3}{2}^{+}}
=M_{\Sigma^{\ast -}}^2+M_{\Omega_{cc}^{\ast +}}^2-2M_{\Xi_c^{\ast 0}}^2
 \label{subeq:2dc-3+} .
\end{aligned}
\end{equation}
%
%

    Considering the isospin breaking effects,
     Eq. (61) in Ref. \cite{Regge2008} can be expressed as
\begin{equation}
\begin{aligned}
(M_{\Omega_{cc}^{+}}^2-M_{\Xi_{cc}^{++}}^2)+(M^2_{\Xi^0}-M^2_{\Sigma^+})=(M_{\Omega_c^0}^2-M_{\Sigma_c^{++}}^2) ,\\  
(M_{\Omega_{cc}^{+}}^2-M_{\Xi_{cc}^{+}}^2)+(M^2_{\Xi^-}-M^2_{\Sigma^-})=(M_{\Omega_c^0}^2-M_{\Sigma_c^0}^2) \label{Omega-cc-d1+} .
\end{aligned}
\end{equation}
    The linear forms in Eq. (\ref{Omega-cc-d1+})    were given
  by Verma and Khanna considering the second-order effects arising from the \underline{84} representation of SU(4) \cite{Verma-Khanna-SU4}
 and by Singh \emph{et al}. studying SU(4) second-order mass-breaking effects with a dynamical consideration \cite{Singh-Verma-Khanna}.
   The linear forms in Eq. (\ref{Omega-cc-d1+})   can satisfy
 the instanton model \cite{instanton-model} and the SU(8) symmetry \cite{SU8-Hendry-Verma}.

    As done in Ref. \cite{Regge2008}, assuming that $\delta_{nc}^{\frac{1}{2}^+}$=$\delta_{nc}^{\frac{3}{2}^+}$,
  inserting the masses of $N^+$, $N^0$, $\Sigma^+$, $\Sigma^-$, $\Xi^0$, $\Xi^-$, $\Lambda_c^+$, $\Sigma_c^{++}$, $\Sigma_c^+$, $\Sigma_c^0$,  $\Omega_c^0$, $\Xi_c^{*+}$, $\Xi_c^{*0}$ from RPP \cite{PDG2014}
 and the expression for $M_{\Omega_{cc}^{*+}}$ obtained in Sec. II
  into Eqs.   (\ref{subeq:uc+dc-1+}), (\ref{subeq:2dc-3+}), and (\ref{Omega-cc-d1+}),  we can have the expressions for  $M_{\Omega_{cc}^+}$,   $M_{\Xi_{cc}^{++}}$, and $M_{\Xi_{cc}^+}$.
    Then, we can get their values:  $M_{\Omega_{cc}^+}$=$3650\pm40$ MeV,   $M_{\Xi_{cc}^{++}}$=$3521^{+41}_{-40}$ MeV, $M_{\Xi_{cc}^+}$=$3520^{+41}_{-40}$ MeV (truncated to the 1 MeV digit).
    We use $M_{\Xi_{cc}^{}}$ to note the averaged mass of  ${\Xi_{cc}^{++}}$ and ${\Xi_{cc}^{+}}$.   Therefore, we can obtain the expression for $M_{\Xi_{cc}^{}}$ and get its  value to be 3520$^{+41}_{-40}$ MeV.
    We can also obtain the expression for the  isospin splitting $M_{\Xi_{cc}^{++}} - M_{\Xi_{cc}^{+}}$  and get its  value to be $0.4 \pm 0.3$ MeV (where the uncertainties come from the errors of the input data).
    In the calculation, we avoid using the masses of $\Delta^{++}$, $\Delta^{+}$, $\Delta^{0}$, and $\Delta^{-}$ because 
 only the charge-mixed states of $\Delta(1232)$ were reliably measured, as indicated in RPP  \cite{PDG2014}.   
    Comparison of the masses of $\Xi_{cc}$ and $\Omega_{cc}^{+}$  extracted in the present paper and those given in other references is shown in Table 4 and Table 5, respectively.

    Then, with the mass expressions for $\Xi_{cc}^{}$, $\Omega_{cc}^{+}$, $\alpha^\prime_{scc}$ and $\alpha^\prime_{ucc}$ obtained above,
   the masses of   the  orbitally excited baryons ($\frac{3}{2}^{-}$, $\frac{5}{2}^{+}$, and $\frac{7}{2}^{-}$) lying on the  $\Xi_{cc}^{}$ and $\Omega_{cc}^{+}$  trajectories  can be expressed as functions of masses of light baryons and singly charmed baryons.
%
    The numerical results are also shown in Table 4 and Table 5.
%


\section{Discussion  and Summary}

    In the present work, we focused on  studying  masses of doubly and triply charmed baryons which do not rely on  unobservable parameters and distrustful resonances.
%
    Under the  Regge phenomenology, we  first expressed  the  mass  of  the ground state ($L=0$) doubly charmed baryon  $\Omega_{cc}^{*+}$
 as a function of masses of light baryons ($\Sigma^{*+}$, $\Xi^{*0}$, $\Omega^-$) and singly charmed baryons ($\Xi_c^{*+}$, $\Omega_{c}^{*0}$) which are well established experimentally \cite{PDG2014}.
    Then, with the quadratic mass relations derived from Regge phenomenology, we calculated the  masses of  the ground state triply charmed baryon $\Omega_{ccc}^{++}$ and  doubly charmed baryons $\Xi_{cc}^{(*)++}$, $\Xi_{cc}^{(*)+}$, and $\Omega_{cc}^{+}$.
%
%
 After that, masses of the orbitally excited (the orbital quantum number $L$=1,2,3) doubly and triply charmed baryons were estimated.    
    In this work, all the input masses of baryons used in the calculation were taken from the Particle Data Group's latest ``Review of Particle Physics''  \cite{PDG2014} 
 and the isospin splittings were also considered.
    The uncertainties of the results only come from the errors of the input data.
  Regge slopes used in this work were also estimated from light and singly charmed baryons.
%
    No systematic error due to any small deviations from the Regge trajectories has been taken into account in this work.

   From Tables 1-5, we can see that the masses of ground state and orbitally excited  doubly and triply charmed baryons predicted here  are reasonable
 comparing with the existing experimental data and those given in many other different approaches. 
    The mass relations and the predictions may be useful for the discovery of the unobserved doubly and triply charmed baryon states and
the $J^P$ assignment of these  baryon states when they are observed in the near future.

    In Ref. \cite{Bjorken:1985ei}, Bjorken  pointed out $\frac{M_{\Omega_{ccc}}}{M_{\psi}}=1.59\pm0.03$ and gave the mass of $\Omega_{ccc}$ to be 4925$\pm$90 MeV,
 which agrees with our result $M_{\Omega_{ccc}}$=4834$^{+82}_{-81}$ MeV shown in Table 2.
   In the present work, the central value of  mass splittings  ($M_{\Omega_{cc}^{*+}}-M_{\Xi_{cc}^{*}} = 3809-3695$ =114 MeV and $M_{\Omega_{cc}^+}-M_{\Xi_{cc}^{}} = 3650-3520$ =130 MeV)  are reasonable.
    The central value of mass splittings  ($M_{\Omega_{cc}^{*+}}-M_{\Omega_{cc}^+} = 3809-3650$ =159 MeV and $M_{\Xi_{cc}^{*}}-M_{\Xi_{cc}^{}} = 3695-3520$ =175 MeV) are a little big.     
    In Ref. \cite{Giannuzzi:2009gh}, $M_{\Xi^*_{cc}}-M_{\Xi_{cc}} = 3719-6547 = 172$ MeV, agrees with our present results.   
    The mass splitting obtained in the framework of nonrelativistic effective field theories of QCD is $M_{\Xi_{cc}^*}-M_{\Xi_{cc}}=120 \pm 40 $ MeV (see Ref. \cite{Brambilla-Vairo-2005} and references therein).
   The isospin splitting  in the present paper,  $M_{\Xi_{cc}^{++}} - M_{\Xi_{cc}^{+}} = 0.4 \pm 0.3$ MeV,   is comparable with  1.5$\pm$2.7 MeV in Ref. \cite{Brodsky:2011zs} and 2.3$\pm$1.7 MeV in Ref.  \cite{Hwang:2008dj}.
    These can be tested by experiments in the future.
    In  Ref. \cite{Brodsky:2011zs} strong and electromagnetic sources of isospin breaking are handled  differently.
    Regge theory  appears to be a pure QCD emergent phenomenon.
  In this work, we do not consider the  electromagnetic corrections separately because the electromagnetic corrections cancel out  in Eqs. (\ref{subeq:3sc-3+}) and (\ref{Omega-cc-d1+}).
 (We would like to thank the anonymous referee for his/her valuable suggestion.)
%

%
    The doubly charmed baryon $\Xi_{cc}^+(3520)$ (ccd) was first reported in the charged weak decay mode $\Xi_{cc}^+ \rightarrow \Lambda_c^+ K^- \pi^+$ (SELEX 2002~\cite{Xi_cc-SELEX2002}),
 with mass $M$=3519$\pm$1 MeV.
    $\Xi_{cc}^+(3520)$ was confirmed in another charged weak  decay mode $\Xi_{cc}^+ \rightarrow p D^+ K^-$ (SELEX 2005~\cite{Xi_cc-SELEX2005}),
  with mass $M$=3518$\pm$3 MeV.
   These reports were adopted by \emph{Particle Data Group} \cite{PDG2014} with the average mass 3518.9$\pm$0.9 MeV.
   However, the $J^P$ number has not been determined experimentally.
   Moreover, it has not been confirmed by other experiments (notably by  LHCb \cite{Xi_cc(3520)-LHCb}, BELLE \cite{Xi_cc(3520)-BELLE}, BABAR \cite{Xi_cc(3520)-BABAR}, and FOCUS \cite{Xi_cc(3520)-FOCUS})
 even though they have  more reconstructed charm baryons than SELEX.

    In the present work, in Sec. IV, the mass of the ${1\over{2}} ^+$ doubly charmed baryon $\Xi_{cc}^+$ (ccd) was predicted to be 3520$^{+41}_{-40}$ MeV.
    In the previous work \cite{Regge2008}, we proved that  the mass of $\Xi_{cc}^+(3520)$ is too small to be assigned as the $\frac{3}{2}^+$ doubly charmed baryon.
    The mass of $\Xi_{cc}^+$ obtained in the present paper agrees with the mass of $\Xi_{cc}^+(3520)$.
    Therefore, we support that the state $\Xi_{cc}^+(3520)$ really exists. We suggest that the $J^P$ of $\Xi_{cc}^+(3520)$ is $\frac{1}{2}^+$. %
   This assignment coincides with the fact that $\Xi_{cc}^+(3520)$ is observed to decay only weakly \cite{Xi_cc-SELEX2002, Xi_cc-SELEX2005}
 (if the $J^P$ of $\Xi_{cc}^+(3520)$ were $\frac{3}{2}^+$, it should decay electromagnetically \cite{Xi_cc(3520)-not-3/2})
 and agrees with many theoretical discussions \cite{Chiu:2005zc, Cohen-Liu-Martynenko-Edwards-Chang, Tang:2011fv, Martynenko:2007je}.  %

   In this work, we took squared mass relations rather than  linear mass relations taken in
 Refs. \cite{{instanton-model},  {Verma-Khanna-SU4}, {SU8-Hendry-Verma}, {Singh-Verma-Khanna}}.
    In the light quark sector, the linear mass relations and the quadratic mass relations lead to the similar results.
    However, they do lead to different values when mass relations include light, charmed, and doubly charmed baryons such as Eqs. (\ref{subeq:3sc-3+}) and (\ref{Omega-cc-d1+}).
    Searching for doubly and triply charmed baryons is helpful not only for the abundance of baryon spectra,
 but also for numerically examining whether the linear mass relations or the quadratic mass relations are realized in nature.
     The triply charmed baryon $\Omega_{ccc}$ is of considerable theoretical interest \cite{Jia-2006,Brambilla:2013vx}.
 %
    Therefore, more efforts should be given to study doubly and triply charmed baryons both theoretically and experimentally.

    The approach presented in the present paper can be used to calculate the masses of the doubly and triply charmed baryons based on  the well established light  and singly charmed baryons  
 while the approach in the previous work \cite{Regge2008} needs the mass of one doubly heavy baryon to predict the masses of other doubly heavy baryons.
    We will estimate masses of doubly and triply bottom baryons in our next work.

 \vspace{-0.1 cm}
\begin{acknowledgments}
  This work was supported in part by National Natural Science Foundation of China (Project Nos. U1204115,  11175020, 11275025, and  11305003).
\end{acknowledgments}


\newpage

\begin{table} [h]  
Table 1.   The masses of the doubly  charmed baryons lying on the $\Omega_{cc}^{*+}$ trajectory  (in units of MeV).   
    Our results are labelled with ``This work" while experimental data from Ref. \cite{PDG2014} are labelled with ``RPP".
    The masses of the ground state $\Omega_{cc}^{*+}$  in other theoretical references vary in the range 3700-3880  MeV
\begin{ruledtabular}
\renewcommand\arraystretch{0.9}
\begin{tabular}{r    *{5}{l}}
 $J^P$                            & $\frac{3}{2}^+$               & $\frac{5}{2}^-$               &$\frac{7}{2}^+$                & $\frac{9}{2}^-$                    \\ \hline                        

This work                           & 3809$\pm$36         & 4058$^{+60}_{-59}$    & 4294$^{+81}_{-79}$    & 4516$^{+100}_{-98}$     \\         

RPP. \cite{PDG2014}                 &&&              \\

Ref. \cite{Roberts:2007ni}           &3876     &4152    &4230    &     &             \\

Ref. \cite{Kiselev-etal-2000-2002}          &3730   &4134    &4204   &   &                        \\

Ref. \cite{Regge2008}                              &3808.4$\pm$4.3                 &                           &4313$\pm$23                        & \\            

Ref. \cite{Ebert:2002-05-2007}         &3872      & 4303      \\

Ref. \cite{Wang:2010vn}                     &3760$\pm$170       &    &       &       \\

Ref. \cite{Chiu:2005zc}                     &3762$\pm$17       &    &       &       \\

Ref.  \cite{H.R.Petry}                        &3765      &    &      &   &             \\
Ref.  \cite{Ponce-1979}                &3764     &     &     &     &                     \\

Ref. \cite{Burakovsky1997}                              &$3850 \pm 25$                            \\


Ref. \cite{Martynenko:2007je}          &3746        \\

Ref. \cite{Bjorken:1985ei}                  &3840$\pm$60                      \\
Ref. \cite{Flynn-etal-2003}                 &3734$\pm14 \pm$8$\pm$97                         \\

Ref. \cite{Roncaglia:1995az}         &3820$\pm$80          \\

Ref. \cite{He:2004px}                                                &3721               \\

Ref. \cite{Martin-Richard-1995}                &3797     &      &                      \\

Ref. \cite{Brown:2014ena}                      &3822$\pm$20$\pm$22        &   &      &     &        \\
Ref. \cite{Brown:2014pra}                      &3773$\pm$38        &   &      &     &              \\
Ref. \cite{Ghalenovi:2014swa}                      &3758       &    &       &       \\
Ref. \cite{Aliev:2012iv}                   &3780$\pm$160       &    &       &       \\
Ref. \cite{Briceno:2012wt}                      &3765$\pm$43$\pm$17$\pm$5       &    &       &       \\

Ref. \cite{Gerasyuta:1999pc}            &3700       &    &       &       \\
Ref. \cite{Alexandrou:2014sha}                         &3735$\pm$33$\pm$18$\pm$43       &    &       &       \\
Ref. \cite{Zhang:2008rt}                               &3810$\pm$60       &    &       &       \\

Ref. \cite{Patel:2008mv}              & 3651-3782       &      &   &      &                           \\

Ref. \cite{Ghalenovi:2011zz}         & 3847       &      &   &      &                           \\
Ref. \cite{Bernotas:2008bu}          & 3800       &      &   &      &                           \\
Ref. \cite{Jaffe:1975us}             & 3795       &      &   &      &                           \\
Ref. \cite{Kiselev:2001fw}            & 3690       &      &   &      &                           \\

Ref. \cite{Tang:2011fv}               & 3710       &      &   &      &                           \\
Ref. \cite{Giannuzzi:2009gh}          & 3770       &      &   &      &                           \\

Ref. \cite{Valcarce:2008dr}          & 3769

\end{tabular}

\end{ruledtabular}
\end{table}

\begin{table} [h]  
Table 2.   The masses of the  triply charmed baryons lying on the $\Omega_{ccc}^{++}$ trajectory  (in units of MeV).     
    Our results are labelled with ``This work" while experimental data from Ref. \cite{PDG2014} are labelled with ``RPP".
    The masses of the ground state $\Omega_{ccc}^{*+}$  in other theoretical references vary in the range 4700-4950  MeV.
\begin{ruledtabular}
\renewcommand\arraystretch{0.9}
\begin{tabular}{r |  *{5}{l}}
 $J^P$                              & $\frac{3}{2}^+$               & $\frac{3}{2}^-$               &$\frac{7}{2}^+$                & $\frac{7}{2}^-$                    \\ \hline        

This work                     & 4834$^{+82}_{-81}$         & 5073$^{+109}_{-107}$    & 5301$^{+134}_{-131}$    & 5520$^{+156}_{-154}$    \\          

RPP. \cite{PDG2014}                  &&& \\

Ref. \cite{Roberts:2007ni}               &4965   & 5160      &5331   &       &\\

Ref.  \cite{Regge2008}                 &4818.9$\pm$6.8   &       &5302$\pm$21    & \\            

Ref. \cite{Wang:2010vn}        &4990$\pm$140       &5110$\pm$100       &       &       \\

Ref. \cite{Chiu:2005zc}                     &4681$\pm$28       &5066$\pm$48    &       &       \\

Ref. \cite{H.R.Petry}            &4773       & 5041                      \\

Ref.  \cite{Ponce-1979}               &4747    &       &    &       &       \\

Ref. \cite{Burakovsky1997}              &4930$\pm$45                       \\

Ref. \cite{Martynenko:2007je}       &4803      \\

Ref. \cite{Bjorken:1985ei}           &4925$\pm$90               \\


Ref. \cite{Martin-Richard-1995}      &4787               \\

Ref.  \cite{Jia-2006}               & 4760$\pm$60               \\
Ref. \cite{Hasenfratz:1980ka}         & 4790               \\

Ref. \cite{Vijande-etal-2004}         &4632               \\

Ref. \cite{Brown:2014ena}            &4796$\pm$8$\pm$18        &    &       &       \\
Ref. \cite{Brown:2014pra}            &4794$\pm$9       &    &       &       \\
Ref. \cite{Ghalenovi:2014swa}        &4880       &    &       &       \\
Ref. \cite{Briceno:2012wt}           &4761$\pm$52$\pm$61$\pm$6       &    &       &       \\

Ref. \cite{Gerasyuta:1999pc}       &4792    &      &   &      &          \\
Ref. \cite{Alexandrou:2012xk}        &4676$\pm$46$\pm$30      &   &      &          \\
Ref. \cite{Alexandrou:2014sha}       &4734$\pm$12$\pm$11$\pm$9    &      &   &      &                      \\
Ref. \cite{Zhang:2008rt}              & 4670$\pm$150       &      &   &      &                           \\

Ref. \cite{Patel:2008mv}              & 4728-4897       &      &   &      &                           \\

Ref. \cite{Ghalenovi:2011zz}         & 4978       &      &   &      &                           \\
Ref. \cite{Bernotas:2008bu}          & 4777       &      &   &      &                           \\
Ref. \cite{Jaffe:1975us}             & 4827       &      &   &      &                           \\

Ref. \cite{Valcarce:2008dr}          & 4758       & 5060     & 5300                               \\

Ref. \cite{Aliev:2014lxa}             & 4720$\pm$120       & 4900$\pm$100      &   &      &                           \\

Ref. \cite{Padmanath:2013zfa}    &4761   & 5123      &5396    &5680 \\            

Ref. \cite{LlanesEstrada:2011kc}      & 4900$\pm$250       &      &   &      &                           \\
Ref. \cite{Flynn:2011gf}             & 4799       &      &   &      &                           \\

\end{tabular}

\end{ruledtabular}
\end{table}

\begin{table} [h]  
Table 3.   The masses of the doubly  charmed baryons lying on the $\Xi_{cc}^{*}$ trajectory  (in units of MeV).     
    Our results are labelled with ``This work" while experimental data from Ref. \cite{PDG2014} are labelled with ``RPP".
    The masses of the ground state $\Xi_{cc}^{*}$ in other theoretical references vary in the range  3600-3750  MeV.
     The  isospin splitting $M_{\Xi_{cc}^{*++}} - M_{\Xi_{cc}^{*+}}$=$1.3^{+1.1}_{-1.2}$ MeV.
\begin{ruledtabular}
\renewcommand\arraystretch{0.9}
\begin{tabular}{r | *{5}{l}  }
 $J^P$                & $\frac{3}{2}^+$              & $\frac{5}{2}^-$       &$\frac{7}{2}^+$       & $\frac{9}{2}^-$      \\ \hline     

This work               & 3695$\pm$34           & 3949$^{+59}_{-58}$        & 4187$^{+81}_{-80}$    & 4413$^{+101}_{-100}$     \\       

RPP. \cite{PDG2014}    &&&                          \\

Ref. \cite{Roberts:2007ni}   &3753     &4092    &4097    &     &                     \\

Ref. \cite{Kiselev-etal-2000-2002}     &3610    &4047    &4089   &     &                              \\

Ref. \cite{Regge2008}    &3684.4$\pm$4.4        &                               &4192$\pm$19                    &  \\                       

Ref. \cite{Ebert:2002-05-2007}       &3727       &4155      &   &      &              \\

Ref. \cite{Wang:2010vn}          &3610$\pm$180    &      &   &      &                 \\

Ref. \cite{Chiu:2005zc}                     &3655$\pm$20       &    &       &       \\

Ref.  \cite{H.R.Petry}        &3723     &    &    &     &                             \\
Ref.  \cite{Ponce-1979}      &3630     &     &    &      &                                  \\

Ref. \cite{Burakovsky1997}                    &$3735 \pm 17$         &      &   &      &                                  \\

Ref. \cite{Martynenko:2007je}         &3548   &      &   &      &       \\

Ref. \cite{Bjorken:1985ei}                   &3695$\pm$60   &      &   &      &                        \\
Ref. \cite{Flynn-etal-2003}               &3641$\pm18 \pm$8$\pm$95      &      &   &      &                        \\

Ref. \cite{Roncaglia:1995az}    &3740$\pm$70     &      &   &      &            \\

Ref. \cite{He:2004px}                     &3630        &      &   &      &                                   \\

Ref. \cite{Vijande-etal-2004}           &3548$\pm$24    &      &   &      &            \\


Ref. \cite{Brown:2014ena}        &3692$\pm$68$\pm$61        &    &       &     &                 \\
Ref. \cite{Brown:2014pra}        &3627$\pm$54       &    &       &     &                        \\
Ref. \cite{Karliner:2014gca}     &3690$\pm$12    &      &   &      &              \\
Ref. \cite{Ghalenovi:2014swa}    &3623    &      &   &      &                     \\
Ref. \cite{Aliev:2012iv}         &3690$\pm$160    &      &   &      &                \\

Ref. \cite{Briceno:2012wt}       &3648$\pm$42$\pm$18$\pm$7    &      &   &      &                      \\

Ref. \cite{Gerasyuta:1999pc}       &3597    &      &   &      &                      \\
Ref. \cite{Alexandrou:2012xk}        &3571$\pm$25      &   &      &                   &      \\
Ref. \cite{Alexandrou:2014sha}       &3652$\pm$17$\pm$27$\pm$3    &      &   &      &                     \\
Ref. \cite{Zhang:2008rt}            &3900$\pm$100    &      &   &      &                       \\
Ref. \cite{Bagan:1992za}           &3580$\pm$50    &      &   &      &                       \\

Ref. \cite{Patel:2008mv}              & 3537-3684        &      &   &      &                           \\

Ref. \cite{Ghalenovi:2011zz}         & 3711       &      &   &      &                           \\
Ref. \cite{Bernotas:2008bu}          & 3661       &      &   &      &                           \\
Ref. \cite{Jaffe:1975us}             & 3661       &      &   &      &                           \\
Ref. \cite{Kiselev:2001fw}            & 3610       &      &   &      &                           \\

Ref. \cite{Tang:2011fv}               & 3620       &      &   &      &                           \\
Ref. \cite{Giannuzzi:2009gh}          & 3719       &      &   &      &                           \\

Ref. \cite{Valcarce:2008dr}          & 3656                             \\

\end{tabular}

\end{ruledtabular}
\end{table}

\begin{table} [h]  
Table 4. The masses of the doubly charmed baryons lying on the $\Xi_{cc}^{}$ trajectory (in units of MeV).   
    Our results are labelled with ``This work" while experimental data from Ref. \cite{PDG2014} are labelled with ``RPP".
    The masses of the ground state $\Xi_{cc}^{}$ in other theoretical references vary in the range 3480-3650  MeV.
  In the previous work~\cite{Regge2008}, the mass of $\Xi_{cc}(3520)$ was  taken as the input value of $\Xi_{cc}^{}$.  
     The  isospin splitting $M_{\Xi_{cc}^{++}} - M_{\Xi_{cc}^{+}}$=$0.4 \pm 0.3$ MeV.
\begin{ruledtabular}
\renewcommand\arraystretch{0.9}
\begin{tabular}{r | lllll  }

 $J^P$                & $\frac{1}{2}^+$               & $\frac{3}{2}^-$               &$\frac{5}{2}^+$                & $\frac{7}{2}^-$                         \\ \hline         

This work            & 3520$^{+41}_{-40}$         & 3786$^{+66}_{-65}$    & 4034$^{+89}_{-87}$    & 4267$^{+109}_{-107}$          \\               

RPP. \cite{PDG2014}            &3518.9$\pm$0.9      &    &                         &   \\   

Ref. \cite{Roberts:2007ni}       &3676   &3921    &4047   &    &                    \\
Ref. \cite{Kiselev-etal-2000-2002}     &3478   &3834   &4050   &    &                     \\

Ref. \cite{Regge2008}    &3518.9$\pm$0.9  &     &4047$\pm$19  &       \\      

Ref. \cite{Ebert:2002-05-2007}   &3620 &3959    &   &    &         \\

Ref. \cite{Wang:2010vn}                 &3570$\pm$140       &    &       &       \\

Ref. \cite{Chiu:2005zc}                     &3522$\pm$16       &    &       &       \\

Ref. \cite{H.R.Petry}        &3642   &3920   & &   &              \\

Ref. \cite{Ponce-1979}       &3511    &    &    &    &         \\

Ref. \cite{Burakovsky1997}              &$3610 \pm 3$       &    &   &    &                                       \\

Ref. \cite{Martynenko:2007je}            &3510  &    &   &    &      \\

Ref. \cite{Bjorken:1985ei}              &3635   &    &   &    &               \\
Ref. \cite{Flynn-etal-2003}          &3549$\pm13 \pm$19$\pm$92   &    &   &    &                               \\

Ref. \cite{Roncaglia:1995az}     &3660$\pm$70   &   &   & &       \\

Ref. \cite{He:2004px}                     &3520      &    &   &    &                               \\


Ref. \cite{Kiselev:2000jb}               &3550$\pm$80        &    &   &    &                      \\

Ref. \cite{Vijande-etal-2004}         &3524   &    &   &    &                  \\

Ref. \cite{Brown:2014ena}        &3610$\pm$23$\pm$22      &    &      &   &                    \\
Ref. \cite{Brown:2014pra}        &3558$\pm$39             &    &      &   &                \\
Ref. \cite{Karliner:2014gca}     &3627$\pm$12    &      &   &      &                   \\
Ref. \cite{Ghalenovi:2014swa}    &3532    &      &   &      &                  \\
Ref. \cite{Briceno:2012wt}       &3595$\pm$39$\pm$20$\pm$7    &      &   &      &                     \\

Ref. \cite{Gerasyuta:1999pc}       &3527    &      &   &      &                       \\
Ref. \cite{Weng:2010rb}            &3520-3560    &      &   &      &                         \\
Ref. \cite{Alexandrou:2012xk}           &3513$\pm$23$\pm$24      &   &      &                   &        \\
Ref. \cite{Alexandrou:2014sha}       &3568$\pm$14$\pm$19$\pm$1    &      &   &      &                     \\

Ref. \cite{Zhang:2008rt}            &4260$\pm$190    &      &   &      &                     \\
Ref. \cite{Kiselev:1999zj}              &3470$\pm$50    &      &   &      &                  \\
Ref. \cite{Bagan:1992za}              &3480$\pm$50    &      &   &      &        \\

Ref. \cite{Patel:2008mv}              & 3468-3604        &      &   &      &                           \\

Ref. \cite{Ghalenovi:2011zz}          & 3582        &      &   &      &                           \\
Ref. \cite{Bernotas:2008bu}           &3557        &      &   &      &                           \\
Ref. \cite{Jaffe:1975us}              &3538       &      &   &      &                           \\
Ref. \cite{Kiselev:2001fw}            & 3480       &      &   &      &                           \\

Ref. \cite{Tang:2011fv}               & 3519       &      &   &      &                           \\
Ref. \cite{Giannuzzi:2009gh}          & 3547       &      &   &      &                           \\

Ref. \cite{Valcarce:2008dr}          & 3579                              \\

\end{tabular}

\end{ruledtabular}
\end{table}

\begin{table} [h]  
Table 5.   The masses of the doubly charmed baryons lying on the $\Omega_{cc}^{+}$ trajectory (in units of MeV).    
    Our results are labelled with ``This work" while experimental data from Ref. \cite{PDG2014} are labelled with ``RPP".
    The masses of the ground state $\Omega_{cc}^{+}$ in other theoretical references vary in the range 3600-3800  MeV.
\begin{ruledtabular}
\renewcommand\arraystretch{0.9}
\begin{tabular}{r    lllll}

 $J^P$                                              & $\frac{1}{2}^+$               & $\frac{3}{2}^-$               &$\frac{5}{2}^+$                & $\frac{7}{2}^-$   \\ \hline        

This work                                            & 3650$\pm$40              & 3910$^{+64}_{-63}$                & 4153$^{+86}_{-84}$            & 4383$^{+105}_{-103}$     \\      

RPP. \cite{PDG2014}                             & & & \\   

Ref. \cite{Roberts:2007ni}         &3815   &4052   &4202   &  &     \\

Ref. \cite{Kiselev-etal-2000-2002} &3594        &3949    &    &    & \\

Ref. \cite{Regge2008}          &3650.4$\pm$6.3  &   &4174$\pm$26  & \\ 

Ref. \cite{Ebert:2002-05-2007}           &3778  &4102   \\

Ref. \cite{Wang:2010vn}                 &3710$\pm$140       &    &       &       \\

Ref. \cite{Chiu:2005zc}                     &3637$\pm$23       &    &       &       \\

Ref. \cite{H.R.Petry}                  &3732  &3986    \\

Ref. \cite{Ponce-1979}          &3664    \\

Ref. \cite{Burakovsky1997}                          &$3804 \pm 8$                                   \\
Ref. \cite{Martynenko:2007je}                  &3719        \\

Ref. \cite{Bjorken:1985ei}                &3800                  \\
Ref. \cite{Flynn-etal-2003}           &3663$\pm11 \pm$17$\pm$95         &                       \\

Ref. \cite{Roncaglia:1995az}       &3740$\pm$80     \\

Ref. \cite{He:2004px}                    &3619                 &       \\

Ref. \cite{Martin-Richard-1995}           &3737                  \\

Ref. \cite{Kiselev:2000jb}                         &3650$\pm$80             \\

Ref. \cite{Brown:2014ena}                  &3738$\pm$20$\pm$20   &    &       &          \\
Ref. \cite{Brown:2014pra}                  &3689$\pm$38    &    &       &       \\
Ref. \cite{Ghalenovi:2014swa}             &3667       &    &       &       \\
Ref. \cite{Briceno:2012wt}               &3679$\pm$40$\pm$17$\pm$5       &    &       &       \\

Ref. \cite{Gerasyuta:1999pc}           &3598      &   &      &          \\
Ref. \cite{Weng:2010rb}                 &3620-3650      &   &      &          \\
Ref. \cite{Alexandrou:2014sha}               &3658$\pm$11$\pm$16$\pm$50       &    &       &       \\
Ref. \cite{Zhang:2008rt}                   &4250$\pm$200       &    &       &       \\

Ref. \cite{Patel:2008mv}              &  3566-3687       &      &   &      &                           \\

Ref. \cite{Ghalenovi:2011zz}          &3718        &      &   &      &                           \\
Ref. \cite{Bernotas:2008bu}           &3710        &      &   &      &                           \\
Ref. \cite{Jaffe:1975us}              &3690       &      &   &      &                           \\
Ref. \cite{Kiselev:2001fw}            & 3590       &      &   &      &                           \\

Ref. \cite{Tang:2011fv}               & 3630       &      &   &      &                           \\
Ref. \cite{Giannuzzi:2009gh}          & 3648       &      &   &      &                           \\

Ref. \cite{Valcarce:2008dr}          & 3697                              \\

\end{tabular}

\end{ruledtabular}
\end{table}

\end{document}